\documentclass[aps,prl,twocolumn,superscriptaddress]{revtex4}
\usepackage{mathrsfs}
\usepackage{amssymb, amsbsy, amsmath, latexsym, dsfont, array, layout,
graphicx,mathrsfs,color,braket,ulem,bm}
\usepackage{amsfonts}
\usepackage{epsfig}
\usepackage{graphics}
\usepackage{bbold}
\usepackage{psfrag}
\usepackage{mathcomp}
\usepackage{subfigure}
\usepackage{verbatim}
\usepackage[colorlinks,citecolor=blue]{hyperref}

\begin{document}

\title{Dynamical topological invariants and reduced rate functions for dynamical quantum phase transitions in two dimensions}
\author{Xingze Qiu}
\affiliation{Key Laboratory of Quantum Information, University of Science and Technology
of China, CAS, Hefei, Anhui, 230026, China}
\affiliation{Synergetic Innovation Center of Quantum Information and Quantum Physics,
University of Science and Technology of China, Hefei, Anhui 230026, China}
\author{Tian-Shu Deng}
\affiliation{Key Laboratory of Quantum Information, University of Science and Technology
of China, CAS, Hefei, Anhui, 230026, China}
\affiliation{Synergetic Innovation Center of Quantum Information and Quantum Physics,
University of Science and Technology of China, Hefei, Anhui 230026, China}
\author{Guang-Can Guo}
\affiliation{Key Laboratory of Quantum Information, University of Science and Technology
of China, CAS, Hefei, Anhui, 230026, China}
\affiliation{Synergetic Innovation Center of Quantum Information and Quantum Physics,
University of Science and Technology of China, Hefei, Anhui 230026, China}
\author{Wei Yi}
\email{wyiz@ustc.edu.cn}
\affiliation{Key Laboratory of Quantum Information, University of Science and Technology
of China, CAS, Hefei, Anhui, 230026, China}
\affiliation{Synergetic Innovation Center of Quantum Information and Quantum Physics,
University of Science and Technology of China, Hefei, Anhui 230026, China}

\begin{abstract}
We show that dynamical quantum phase transitions (DQPTs) in the quench dynamics of two-dimensional topological systems can be characterized by a dynamical topological invariant defined along an appropriately chosen closed contour in momentum space. Such a dynamical topological invariant reflects the vorticity of dynamical vortices responsible for the DQPTs, and thus serves as a dynamical topological order parameter in two dimensions. We demonstrate that when the contour crosses topologically protected fixed points in the quench dynamics, an intimate connection can be established between the dynamical topological order parameter in two dimensions and those in one dimension. We further define a reduced rate function of the Loschmidt echo on the contour, which features non-analyticities at critical times and is sufficient to characterize DQPTs in two dimensions. We illustrate our results using the Haldane honeycomb model and the quantum anomalous Hall model as concrete examples, both of which have been experimentally realized using cold atoms.
\end{abstract}

\maketitle

The discovery of topological matter extends our understanding of quantum phases and phase transitions beyond the conventional Landau paradigm, where phases of matter are characterized by symmetry properties and local order parameters~\cite{landau}. Instead, topological phases are parameterized by non-local topological invariants, which are related to the topology of ground-state wavefunctions~\cite{HKrmp10,QZrmp11}.
Recent studies of topological phenomena in dynamical processes advance the frontier even further and raise the interesting question on the relation between topology and out-of-equilibrium dynamics~\cite{Demler10,Zoller11,Zollerdiss,Levin13,Gogolin15,Bhaseen15,rigol, Lindner16,Vishwanath16,Sondhi16,Heyl13,Heyl15,Heyl17,Dora15,BH16,Balatsky,Budich16,Refael16,Bhaseen16,Sondhi17,Zhai17,Chen17,Ueda17,
ETHcoldatom14,Jo17,iondtop,Weitenberg17,Weitenberg1709,Xiong-Jun,shuaichen}. An outstanding achievement here is the experimental observation of dynamical quantum phase transitions (DQPTs) in the quench dynamics of a two-dimensional topological system~\cite{Weitenberg17}. The emergence of DQPTs therein are not only intimately connected with the ground-state topology of the initial and final Hamiltonians of the quench, but are also accompanied by the creation or annihilation of dynamical vortices in momentum space.

Proposed by Heyl {\it et al.}~\cite{Heyl13,Heyl17}, the DQPT occurs as the Loschmidt amplitude
$\mathcal{G}(t)=\langle \psi^\text{i}|e^{-iH^\text{f} t}|\psi^\text{i}\rangle$ vanishes at critical times in a quench process, where a system prepared in an eigenstate $|\psi^\text{i}\rangle$ of the initial Hamiltonian $H^\text{i}$ evolves under a distinct final Hamiltonian $H^\text{f}$.
Due to the formal similarity between the Loschmidt amplitude and the canonical partition function with an imaginary temperature, zeros of the Loschmidt amplitude in real time are identified as dynamical Fisher zeros and give rise to dynamical phase transitions~\cite{Weitenberg17,loschmidt}, just as Fisher zeros (or Lee-Yang zeros) in the partition function leading to phase transitions at equilibrium~\cite{fisherbook,LY52}. Correspondingly, at critical times of the DQPT, the rate function of the Loschmidt echo, $g(t)=-1/N \ln |\mathcal{G}(t)|^2$ ($N$ is the number of degrees of freedom)~\cite{Heyl13}, becomes non-analytical in the thermodynamic limit $N\rightarrow\infty$, analogous to the non-analyticity of free energies at thermal critical points.

In one dimension, it has been shown that the onset of DQPTs is characterized by a dynamical topological order parameter (DTOP), which is quantized and can only change its value at critical times~\cite{BH16}. In two dimensions, whereas DQPTs are shown to exist in the quench dynamics of two-dimensional topological systems~\cite{kehrein15,dutta17}, an important question remains regarding the underlying DTOP.
In a recent theoretical study, a generalized winding number has been considered to serve as the dynamical topological invariant~\cite{dutta17ti}. However, its relation with the dynamical vortices, which serve as the effective order parameter in the experiment, is not immediately clear.

Motivated by the experimental observation, we study DQPTs and the underlying DTOP in the quench dynamics of a general two-band topological system in two dimensions. We show how a dynamical topological invariant can be systematically constructed from the Pancharatnam geometric phase (PGP) on closed contours in momentum space, which serves as the DTOP for DQPTs in two dimensions. Interestingly, DTOP in two dimensions are intimately connected to that in one dimension for contours pinned by a pair of topologically protected fixed points, where the PGP vanishes at all times. These fixed points necessarily exist when the system is quenched between topological phases with different absolute value of the Chern number.
Based on these understandings, we define a reduced rate function on the one-dimensional contour, which shows non-analyticities at critical times. This is in contrast to previous discussions, where the rate function is integrated over the two-dimensional Brillouin zone (BZ) and the non-analyticities emerge only in the first derivative of the rate function~\cite{Weitenberg17,higher,kehrein15,dutta17,dutta17ti}. Finally, we illustrate our main results using the Haldane honeycomb model and the quantum anomalous Hall model as two concrete examples. Our work offers a systematic scheme to characterize the DQPT and construct the DTOP in two dimensions, which should greatly simplify their experimental detection.

{\it DQPTs in two dimensions:---}
We consider the quench dynamics of a general two-band model in two dimensions. The Bloch Hamiltonian can be written as
\begin{equation}
H(\mathbf{k})= \mathbf{h}(\mathbf{k})\cdot\bm{\sigma},\label{eqn:Hk}
\end{equation}
where $\bm{\sigma}=(\sigma_1,\sigma_2,\sigma_3)$ is a vector of Pauli matrices. At $t=0$, the system in each quasi-momentum sector is prepared in the ground state $|\psi^\text{i}_-\rangle$ of the initial Hamiltonian $H^\text{i}_{\mathbf k} $, which is then subject to a unitary time evolution governed by the final Hamiltonian $H^\text{f}_{\mathbf k}$. Note that due to the lattice-translational symmetry, the time evolutions in different $\mathbf{k}$-sectors are decoupled.
The state of the system evolves according to $\ket{\psi(\mathbf{k},t)}=\sum_{\alpha=\pm} c_\alpha e^{-i\epsilon^\text{f}_\alpha t}|\psi^\text{f}_\alpha\rangle$, where $c_\alpha:=\langle\psi^\text{f}_\alpha|\psi^{i}_-\rangle$ and $\sum_\alpha|c_\alpha|^2=1$.
Here, $\epsilon^\text{f}_\alpha=\alpha E^\text{f}_{\mathbf{k}}$ and $\ket{\psi^\text{f}_\alpha}$
are respectively the eigenvalues and eigenvectors of $H^\text{f}_{\mathbf k}$.

The Loschmidt amplitude is then
\begin{align}
\mathcal{G}(t)=\prod_{\mathbf{k}\in \text{1BZ}}\mathcal{G}_{\mathbf{k}}(t)=\prod_{\mathbf{k}\in \text{1BZ}}\langle\psi^\text{i}_-(\mathbf{k})|\psi(\mathbf{k},t)\rangle,
\end{align}
where $\mathbf{k}$ runs over the 1BZ. It is then straightforward to derive $\mathcal{G}_{\mathbf{k}}(t):=\left|\mathcal{G}_{\mathbf{k}}(t)\right|e^{i\phi_{\mathbf{k}}(t)} =\left|c_-(\mathbf{k})\right|^{2}e^{iE^\text{f}_{\mathbf{k}}t}+\left|c_+(\mathbf{k})\right|^{2}e^{-iE^\text{f}_{\mathbf{k}}t}$. For future reference, we also define the PGP as $\phi^G_{\mathbf{k}}(t):=\phi_{\mathbf{k}}(t)-\phi^D_{\mathbf{k}}(t)$, where the dynamical phase $\phi^D_{\mathbf{k}}(t)=-\int^t_0dt'\langle\psi(\mathbf{k},t')|H^\text{f}_{\mathbf{k}}\ket{\psi(\mathbf{k},t')}=(|c_-(\mathbf{k})|^2-|c_+(\mathbf{k})|^2)E^\text{f}_{\mathbf{k}}t$.

When either $c_-(\mathbf{k})=0$ or $c_+(\mathbf{k})=0$, $\phi^G_{\mathbf{k}}(t)$ vanishes at all times. We therefore identify the momenta satisfying $c_-=0$ or $c_+=0$ as two distinct kinds of fixed points in the quench dynamics.
Note that $c_\pm$ cannot vanish simultaneously at the same $\mathbf{k}$, as otherwise $|\psi^i_-(\mathbf{k})\rangle=\sum_\alpha c_\alpha(\mathbf{k})|\psi^f_\alpha(\mathbf{k})\rangle=0$. These fixed points correspond to cores of static vortices observed in the azimuthal phase of the time-evolved state~\cite{Weitenberg17}.
Importantly, it has been established that the number of fixed points with $c_\alpha=0$ should be at least $\left|C^\text{i}_- -C^\text{f}_{\alpha}\right|$ each~\cite{sun16,Balatsky,Dora15}, where $C^\beta_{\alpha}$ ($\beta=\text{i},\text{f}$) is the Chern number corresponding to the $\alpha$ band of $H^\beta$. In this sense, the existence of fixed points in the quench dynamics are topologically protected.

Given a pair of fixed points of different kinds, which necessarily exist when the system is quenched between topological phases with different absolute value of the Chern number, we can always connect them by a smooth curve in the BZ. As $|c_+(\mathbf{k})|-|c_-(\mathbf{k})|$ is a continuous function, there must be at least one critical momentum $\mathbf{k}_c$ on the curve, with $|c_-(\mathbf{k}_c)|=|c_+(\mathbf{k}_c)|$.
These critical points should form a closed loop $l_c$ in momentum space, encircling one of the fixed points.
For any given $\mathbf{k}_c\in l_c$, $\mathcal{G}_{\mathbf{k}_c}(t)$ vanishes at critical times $t_c=(2n+1)\pi/2E^\text{f}_{\mathbf{k}_c}$ ($n\in\mathbb{N}$), when $\phi^G_{\mathbf{k}_c}(t_c)$ become ill-defined, corresponding to the emergence of dynamical vortices centered at $\mathbf{k}_c$.
These vortex cores are dynamical Fisher zeros, which emerge, evolve, and annihilates in time on $l_c$. DQPTs occur at times when dynamical Fisher zeros appear on or disappear from $l_c$~\cite{Weitenberg17}.

We note that the existence of $l_c$ and hence DQPTs are topologically protected when they are related to a pair of topologically protected fixed points. Whereas there could be coincidental critical points $\mathbf{k}_c$ and $t_c$ that are not related to fixed points, we focus on the case where only topologically protected critical points exist. Under this condition, dynamical Fisher zeros must appear or disappear in pairs on $l_c$, corresponding to the vortex-anti-vortex pairs observed in the recent experiment~\cite{Weitenberg17}. This is guaranteed by the Hopf theorem, which states that the total vorticity of the BZ should be zero, due to the $T^2$ topology of BZ~\cite{hopf}.

\begin{figure}
\includegraphics[width=6cm]{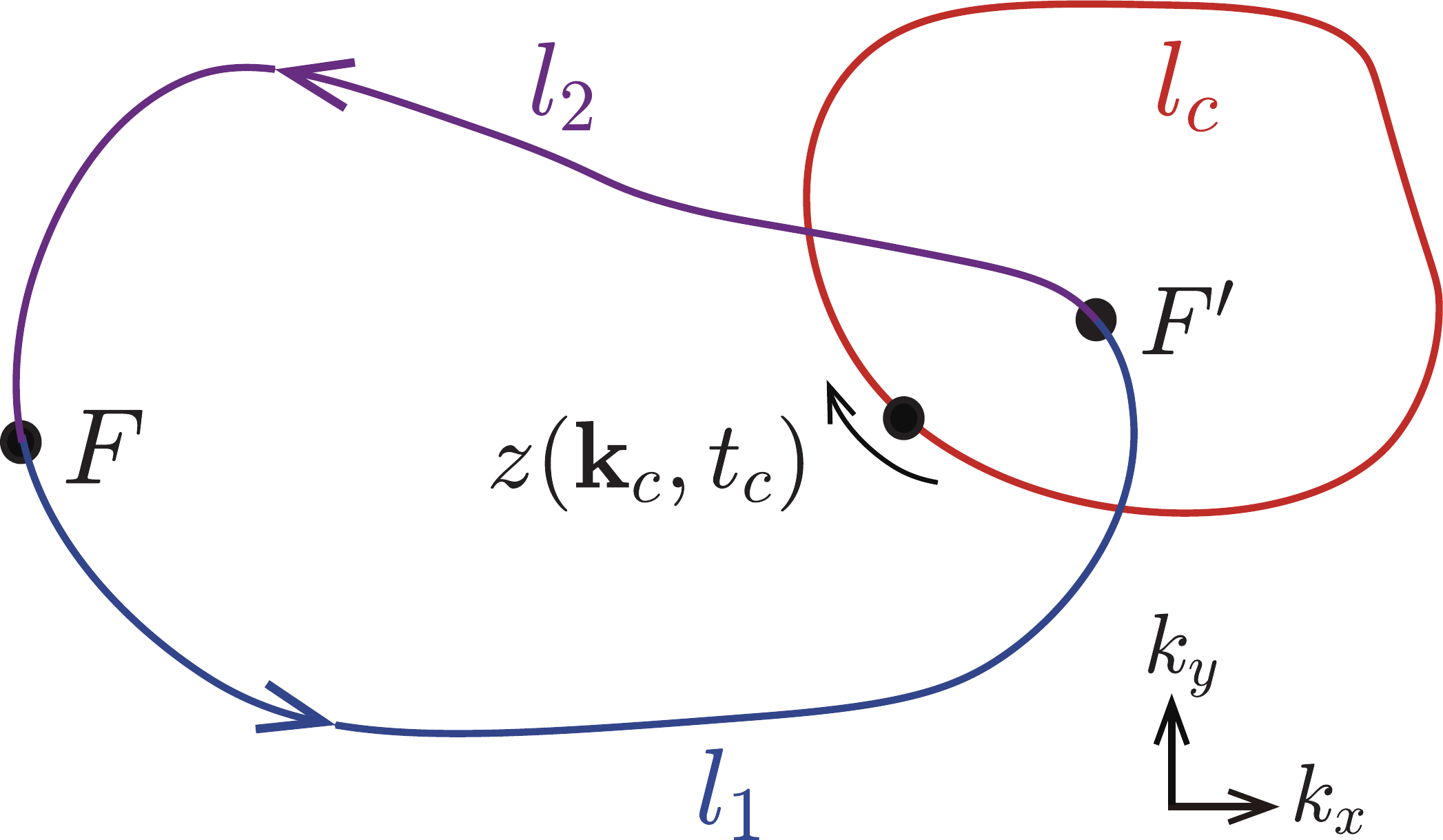}
\caption{(a) $F$ and $F'$ are a pair of distinct fixed points, where either $c_-=0$ or $c_+=0$. On the closed loop $l_c$ (red), we have $|c_-|=|c_+|$, and $z(\mathbf{k}_c,t_c)$ indicates a dynamical Fisher zero, which emerges, evolves and annihilates on $l_c$ during the dynamics. The DTOP $\nu^D$ is defined along the contour $l_w=l_1\bigcup l_2$, where $l_1$ (blue) and $l_2$ (purple) are smooth curves connecting the fixed points. The arrows along $l_w$ indicate the direction of integral.}
\label{fig:fig1}
\end{figure}

{\it DTOP and the reduced rate function:---}
As DQPTs in two dimensions are signaled by dynamical vortices in momentum space, a natural choice of the DTOP is the vorticity of these vortices. To this end, we first define a dynamical winding number on an arbitrary closed contour $l_w$ in the BZ
\begin{equation}
\nu^D(t) := \frac{1}{2\pi}\oint_{l_w} \nabla\phi^G_{\mathbf{k}}(t)\cdot d\bm{l}=\frac{1}{2\pi i}\oint_{l_w}\frac{\partial\mathcal{G}_{\mathbf{k}}(t)}{\mathcal{G}_{\mathbf{k}}(t)}. \label{eqn:dtop}
\end{equation}
Defining the area enclosed by $l_w$ as $S_w$, we have that
$\nu^D(t)$ is the sum of residues of $1/\mathcal{G}_{\mathbf{k}}(t)$ at any given time $t$ in $S_w$, which is essentially the total vorticity of dynamical vortices in $S_w$.

To reflect DQPTs in the overall quench dynamics, one therefore needs to find a contour that encompasses all independent dynamical vortices in the BZ. Note that due to the lattice symmetry as well as the discrete symmetry in vortex-anti-vortex pairs, not all dynamical vortices in the BZ are independent. A systematic way for constructing such a contour is illustrated in Fig.~\ref{fig:fig1}. We start by finding two fixed points of different kinds in the BZ. We then connect them with two smooth curves $l_1$ and $l_2$, both of which intersect with $l_c$. We require these intersections to be the closest points on $l_c$ which have the largest and the smallest $E^{\text{f}}_{\mathbf k}$, respectively. The two intersections thus correspond to locations where the dynamical Fisher zeros appear and disappear, respectively.
Without loss of generality, we assume that the intersection between $l_1$ ($l_2$) and $l_c$ to have the largest (smallest) $E^{\text{f}}_{\mathbf k}$.
Finally, we denote $l_w=l_1\bigcup l_2$, and define the winding number $\nu^D(t)$ according to Eq.~(\ref{eqn:dtop}), which serves as the DTOP for DQPTs in two dimensions.

Interestingly, as the PGP vanishes at fixed points, we can further define winding numbers along $l_1$ and $l_2$, respectively,
\begin{equation}
\nu^D_{1,2}(t) := \frac{1}{2\pi}\int_{l_{1,2}}\nabla\phi^G_{\mathbf{k}}(t)\cdot d\bm{l},
\end{equation}
which characterizes the $S^1\rightarrow S^1$ mapping from $l_1$ ($l_2$) to $e^{i\phi^G_{\mathbf{k}}(t)}$ on $l_1$ ($l_2$). The winding numbers $\nu^D_{1,2}$ can therefore be regarded as DTOPs of ancillary one-dimensional systems along $l_{1,2}$, with the Loschmidt amplitudes
\begin{align}
\tilde{\mathcal{G}}_{1,2}(t)=\prod_{\mathbf{k}\in l_{1,2}}\mathcal{G}_{\mathbf{k}}(t).
\end{align}
According to the theory of DQPT in one dimension~\cite{BH16}, $\nu^D_{1,2}(t)$ are quantized and can only change their values when $\tilde{\mathcal{G}}_{1,2}(t)$ vanish at some critical times. This corresponds to dynamical vortices appearing or disappearing on the boundary at $l_1$ or $l_2$ while entering or leaving $S_w$. Naturally, $\nu^D(t)=\nu^D_{1}(t)+\nu^D_{2}(t)$, i.e., the vorticity of the dynamical vortices in $S_w$ can only change when dynamical vortices move across its boundary.
Therefore, we have shown that DQPTs in the quench process of a two-dimensional system can be characterized by a DTOP defined on an appropriately chosen contour in the BZ, which can be further decomposed into DTOPs along the one-dimensional segments of the contour.

Based on the understanding that information on the appropriately chosen contour $l_w$ is sufficient to characterize DQPTs in two dimensions, we now introduce the reduced rate function
\begin{equation}
g_r(t):=-\frac{1}{2\pi}\int_{\mathbf{k}\in l_w}d\mathbf{k}\ln|\mathcal{G}_{\mathbf{k}}(t)|^2,
\end{equation}
where only the Loschmidt amplitude on the contour $l_w$ contributes. From our previous analysis, $g_r(t)$ demonstrates non-analyticities at critical times of the DQPTs of the two-dimensional system.
We note that, in general, the choice of $l_w$ is not unique, and it does not have to cross the fixed points. As long as $S_w$ contains all the independent dynamical vortices in the BZ, a reduced rate function can be defined on its boundary $l_w$ to characterize DQPTs of the two-dimensional system. In practice, however, as fixed points are typically located at high-symmetry points of the BZ, including them in $l_w$ is intuitive and straightforward.

\begin{figure}
\includegraphics[width=8cm]{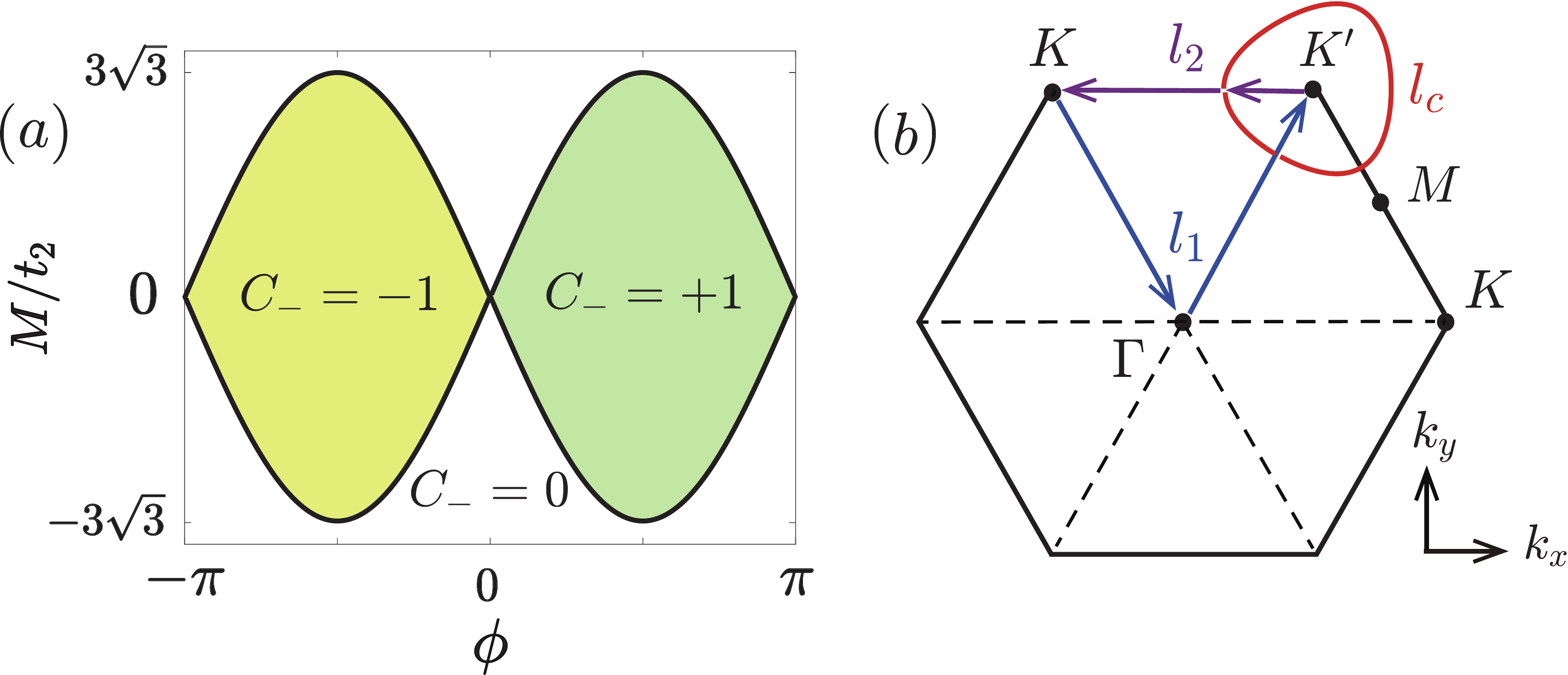}
\caption{(a) Topological phase diagram and (b) the BZ of the Haldane honeycomb model.
For the quench process considered here, the fixed points are located at $K$ and $K'$, where $c_+=0$ and $c_-=0$, respectively. On the closed loop $l_c$ (red), we have $|c_-|=|c_+|$. We define the winding numbers $\nu^D_1$ and $\nu^D_2$ along the loops $l_1$ (blue) and $l_2$ (purple), respectively.}
\label{fig:fig2}
\end{figure}

{\it Haldane honeycomb model:---}
We now study the DQPTs in the quench dynamics of a Haldane honeycomb model as a concrete example~\cite{haldane}.
Under the tight-binding approximation and focusing on the lowest bands, we write the Bloch Hamiltonian $H_{\mathbf{k}}$ in the form of Eq.~(\ref{eqn:Hk}) by neglecting the term proportional to the identity matrix. We have $\mathbf{h}(\mathbf{k})=(h_1,h_2,h_3)$, where $h_1(\mathbf{k})= t_1\sum^3_{j=1}\cos(\mathbf{k}\cdot\mathbf{a}_j)$, $h_2(\mathbf{k})=t_1\sum^3_{j=1}\sin(\mathbf{k}\cdot\mathbf{a}_j)$, and $h_3(\mathbf{k})=M-2t_2\sin\phi\sum^3_{j=1}\sin(\mathbf{k}\cdot\mathbf{b}_j)$. Here, $t_1$ and $t_2$ are respectively the nearest- and next-nearest-neighbor hopping rate, and $M$ is the staggered mass. We also have $\mathbf{a}_1=a_0(\sqrt{3},1)/2$, $\mathbf{a}_2=a_0(-\sqrt{3},1)/2$, $\mathbf{a}_3=a_0(0,-1)$,  $\mathbf{b}_1=\mathbf{a}_2-\mathbf{a}_3$, $\mathbf{b}_2=\mathbf{a}_3-\mathbf{a}_1$, and $\mathbf{b}_3=\mathbf{a}_1-\mathbf{a}_2$, with $a_0$ being the lattice constant.

\begin{figure*}
\includegraphics[width=16cm]{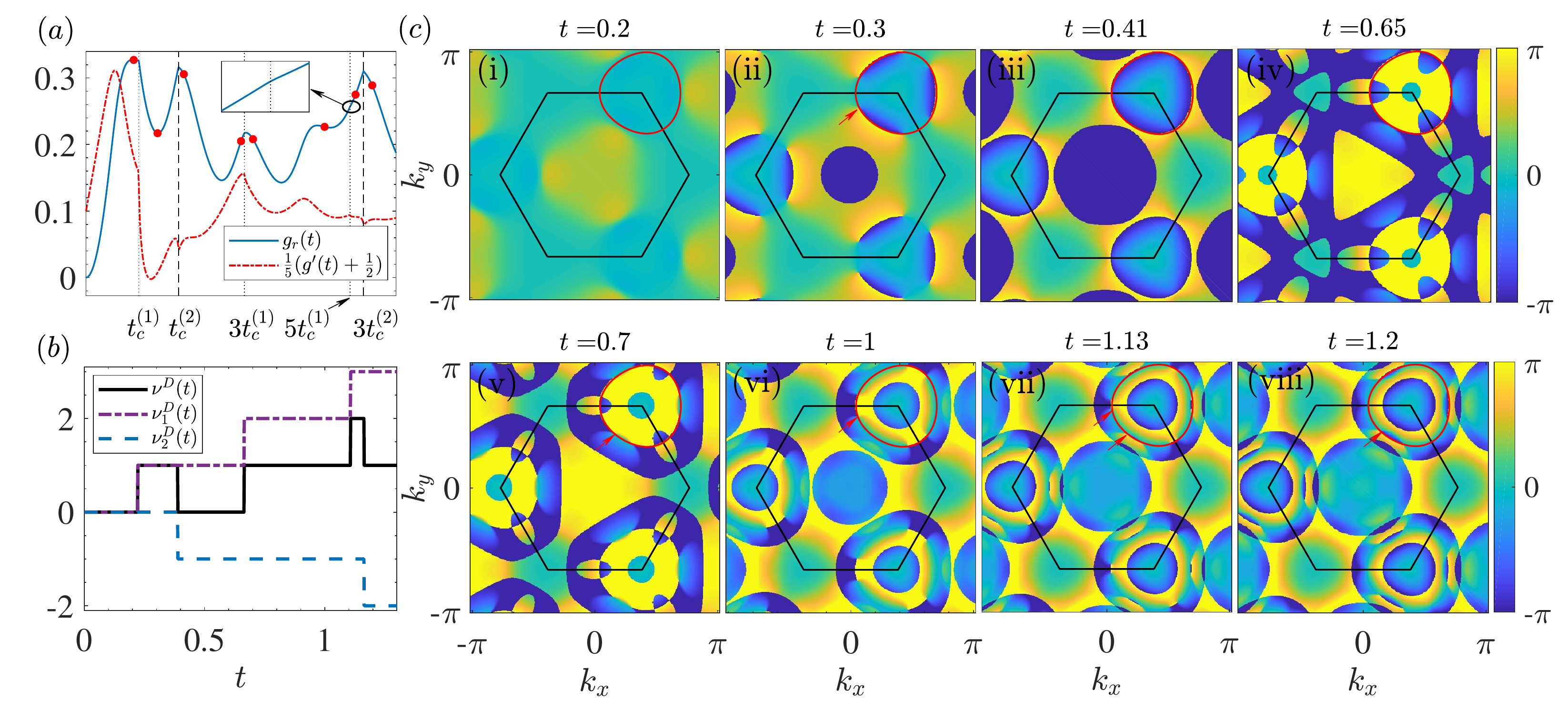}
\caption{DQPT in the quench dynamics of the Haldane honeycomb model. (a) The reduced rate function $g_r(t)$ (blue solid) and the first derivative of the full rate function $g'(t)$ (red dash-dotted). Red dots indicate different times shown in (c), and vertical dash-dotted (dashed) lines indicate the time of appearance (disappearance) of dynamical vortices.
(b) The DTOP $\nu^D(t)$ and the winding numbers $\nu_1^D(t)$, $\nu^D_2(t)$. (c) The phase profile $\phi^G_\mathbf{k}(t)$. Red arrows indicate the cores of dynamical vortices in $S_w$, which evolve in time on $l_c$ (red circle).
Parameters of the quench process are given in the main text, under which the two critical time scales $t_c^{(1)}=\min_{\mathbf{k}\in l_c}\left(\pi/2E^\text{f}_{\mathbf{k}}\right)\approx 0.22$ and $t_c^{(2)}=\max_{\mathbf{k}\in l_c}\left(\pi/2E^\text{f}_{\mathbf{k}}\right)\approx 0.39$. Here the unit of time is taken to be $1/t_2$.}
\label{fig:fig3}
\end{figure*}

We consider a sudden quench process, where the ground state of $H^\text{i}$ evolves according to $H^\text{f}$. For concreteness, we consider the case where $H^\text{i}$ is characterized by the parameters $(t^\text{i}_1/t_2=4,M^\text{i}/t_2=10\sqrt{3},\phi^\text{i}=0)$ and $H^\text{f}$ by $(t^\text{f}_1/t_2=4,M^\text{f}=0,\phi^\text{f}=\pi/2)$. Here we have used $t_2=t^\text{i}_2=t^\text{f}_2$ as the unit of energy. According to the phase diagram Fig.~\ref{fig:fig2}(a), the system is initially in the topologically trivial regime with $C^\text{i}_-=0$, and is quenched into the topologically non-trivial regime with $C^\text{f}_{\pm}=\mp1$. As indicated in Fig.~\ref{fig:fig2}(b), the fixed points in this case are at $K$ and $K'$ of the BZ, with $l_c$ encircling $K'$.
Under the symmetry of the Bloch Hamiltonian, we construct the contour $l_w$ from $l_1$ and $l_2$ as shown in Fig.~\ref{fig:fig2}(b). The segment of $l_c$ intersected by $l_w$ is only one-sixth of $l_c$ but is sufficient to capture the DQPTs of the quench process.

In Fig.~\ref{fig:fig3}(a)(b), we show the reduced rate function $g_r(t)$ calculated on $l_w$, as well as the winding numbers $\nu^D_{1,2}(t)$ and the DTOP $\nu^D(t)$. $g_r(t)$ exhibits non-analyticities at exactly the same critical times as calculated from the derivative of the full rate function $g'(t)=\partial g(t)/\partial t$. Here $g(t)$ is calculated by integrating $-1/(2\pi)^2\ln |\mathcal{G}(t)|^2$ over one-third of BZ~\cite{Weitenberg17} due to the lattice symmetry. In Fig.~\ref{fig:fig3}(a), the critical time scale $t^{(1)}_c$ corresponds to a dynamical vortex emerging at the intersection between $l_1$ and $l_c$, which will enter the region enclosed by $l_w$ immediately afterward. In contrast, the critical time scale $t^{(2)}_c$ corresponds to a dynamical vortex disappearing at the intersection between $l_2$ and $l_c$. These are well captured by $\nu^D(t)$ as shown in Fig.~\ref{fig:fig3}(b), where it directly reflects the total vorticity of dynamical vortices in $S_w$. Interestingly, $\nu^D(t)$ becomes $2$ for $5t^{(1)}_c<t<3t^{(2)}_c$, which suggests the existence of two dynamical vortices in $S_w$. Finally, we plot the time-dependent phase profiles in Fig.~\ref{fig:fig3}(c), where the evolution of the dynamical vortices are clearly visible.

{\it Quantum anomalous Hall model:---}
As another example, we consider the quench dynamics of quantum anomalous Hall model, which has been experimentally realized on an optical Raman lattice~\cite{qah1,qah2,shuaichen}.
Focusing on the lowest $s$-band physics and under the tight-binding approximation, we write the Bloch Hamiltonian in the form of Eq.~(\ref{eqn:Hk}), with
$h_1(\mathbf{k})=2t_{\mathrm{so}}\sin k_y$, $h_2(\mathbf{k})=2t_{\mathrm{so}}\sin k_x$, and $h_3(\mathbf{k})= m_z-2t_0\cos k_x-2t_0\cos k_y$. Here $t_{\mathrm{so}}$ is the Raman-assisted spin-flipping hopping rate, $t_0$ is the nearest-neighbor spin-conserving hopping rate, and $m_z$ is an effective Zeeman field. The ground state of the system at half-filling is topologically non-trivial with a Chern number $C=\text{sgn}(m_z)$ for $|m_z|\in (0,4t_0)$.

\begin{figure}
\includegraphics[width=8cm]{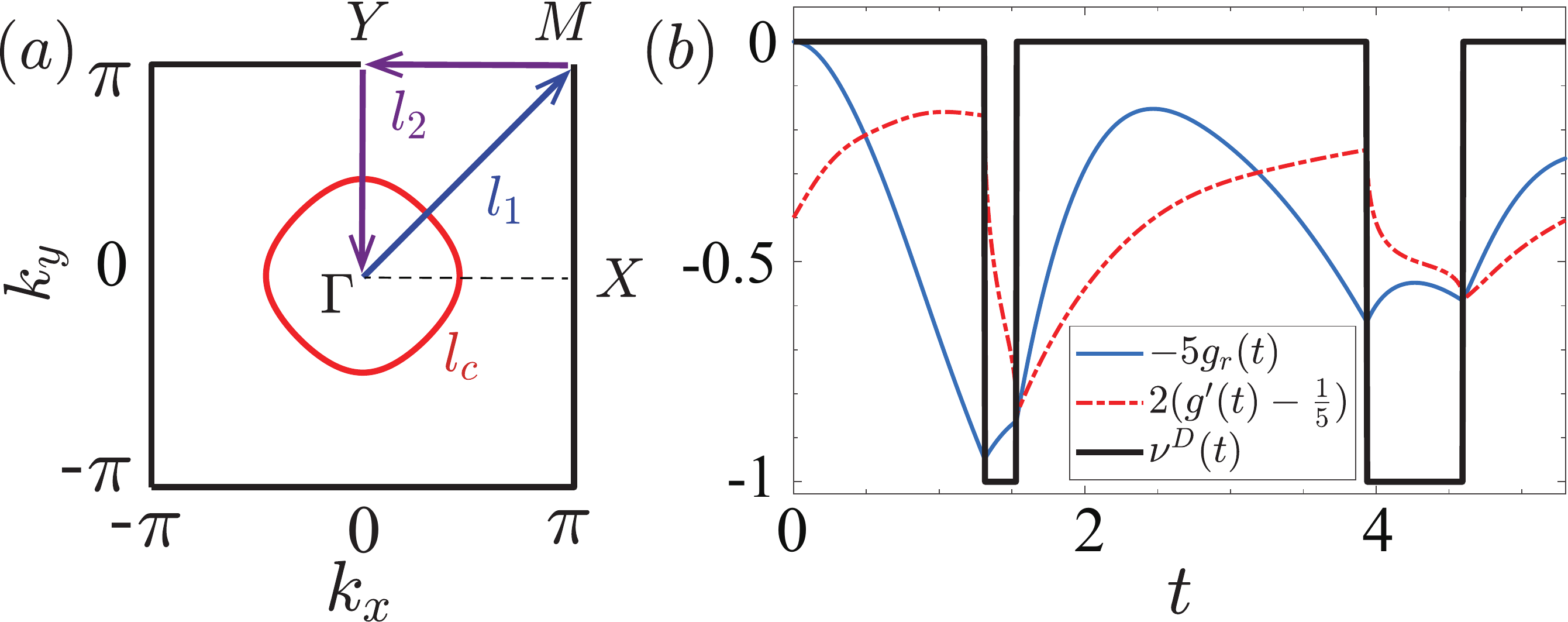}
\caption{DQPTs in the quench dynamics of the Quantum anomalous Hall model. (a) BZ with fixed points and contours. (b) The reduced rate function $g_r(t)$ (blue solid), the DTOP $\nu^{D}(t)$ (black solid), and the first derivative of the full rate function $g'(t)$ (red dash-dotted). Here the Hamiltonian is quenched from $m^\text{i}_z/t_0=6$ to $m^\text{f}_z/t_0=2$, while we fix $t_\mathrm{so}/t_0=1/2$. The two critical time scales are $1.31$ and $1.53$, respectively. The unit of time is taken to be $1/t_0$.}
\label{fig:fig4}
\end{figure}

When the system is quenched from a topologically trivial regime into a topologically non-trivial regime, a pair of topologically protected fixed points exist, located at $\Gamma$ ($c_-=0$) and $M$ ($c_+=0$) points of the BZ, with $l_c$ encircling the $\Gamma$ point.
As illustrated in Fig.~\ref{fig:fig4}, we construct $l_{1,2}$, which encircle one-eighth of $l_c$. We show in Fig.~\ref{fig:fig4} $\nu^{D}(t)$ and $g_r(t)$, which reveal two critical time scales, consistent with results from the derivative of the full rate function. Here the full rate function $g(t)$ is calculated by integrating over one-fourth of the BZ. Note that an additional fixed point exist at $Y$ with $c_+=0$. As we have discussed previously, the inclusion of $Y$ in $l_w$ is convenient but not crucial, as long as the segment of $l_c$ in $S_w$ (see Fig.~\ref{fig:fig4}(a)) is unchanged.

{\it Final remarks:---}
Whereas we mainly focus on the construction of the contour in the presence of a single pair of topologically protected fixed points, our discussion can be seen as the basic building block when more fixed points exist. This is the case, for example, when the quench is between topological phases with a large Chern-number difference. One should then construct a contour for each pair of fixed points, and the overall DTOP, as well as the reduced rate function, should include contributions from all such contours.
We expect that our construction can be extended to higher dimensions, where DQPTs in the quench dynamics of a $d$-dimensional topological system should be captured by a reduced rate function and a DTOP defined in $(d-1)$ dimensions. Our results thus not only greatly simply experimental detection of DQPTs in two dimensions, but also paves the way toward systematic understanding and characterization of DQPTs in higher dimensions.

\begin{acknowledgments}
\textit{Acknowledgement:--}
This work has been supported by the National Natural Science Foundation of China (Grant No. 11522545), the National Key Research and Development Program of China (Grant Nos. 2016YFA0301700, 2017YFA0304100).
\end{acknowledgments}


\begin{thebibliography}{99}
\bibitem{landau} L. D. Landau, Zur Theorie der Phasenumwandlungen. Phys. Z. Sowjetunion {\bf 11}, 26 (1937)
\bibitem{HKrmp10} M. Z. Hasan and C. L.  Kane, Rev. Mod. Phys. {\bf 82}, 3045 (2010).
\bibitem{QZrmp11} X. L. Qi and S. C. Zhang, Rev. Mod. Phys. {\bf 83}, 1057 (2011).

\bibitem{Demler10} T. Kitagawa, E. Berg, M. Rudner, and E. Demler, Phys. Rev. B {\bf 82}, 235114 (2010).
\bibitem{Zoller11} L. Jiang, T. Kitagawa, J. Alicea, A. R. Akhmerov, D. Pekker, G. Refael, J. I. Cirac, E. Demler, M. D. Lukin, and P. Zoller, Phys. Rev. Lett. {\bf 106}, 220402 (2011).
\bibitem{Zollerdiss} S. Diehl, E. Rico, M. A. Baranov, and P. Zoller, Nat. Phys. {\bf 7}, 971 (2011).
\bibitem{Levin13} M. S. Rudner, N. H. Lindner, E. Berg, and M. Levin, Phys. Rev. X {\bf 3}, 031005 (2013).
\bibitem{Gogolin15} J. Eisert, M. Friesdorf, and C. Gogolin, Nat. Phys. {\bf 11}, 124 (2015).
\bibitem{Bhaseen15} M. D. Caio, N. R. Cooper, and M. J. Bhaseen, Phys. Rev. Lett. {\bf 115}, 236403 (2015).
\bibitem{rigol} L. D'Alessio and M. Rigol, Nat. Commun. {\bf 6}, 8336 (2015).
\bibitem{Lindner16} P. Titum, E. Berg, M. S. Rudner, G. Refael, and N. H. Lindner, Phys. Rev. X {\bf 6}, 021013 (2016).
\bibitem{Vishwanath16} A. C. Potter, T. Morimoto, and A. Vishwanath, Phys. Rev. X {\bf 6}, 041001 (2016).
\bibitem{Sondhi16} V. Khemani, A. Lazarides, R. Moessner, and S. L. Sondhi, Phys. Rev. Lett. {\bf 116}, 250401 (2016).
\bibitem{Heyl13} M. Heyl, A. Polkovnikov, and S. Kehrein, Phys. Rev. Lett. {\bf 110}, 135704 (2013).
\bibitem{Heyl17} M. Heyl, arXiv:1709.07461.
\bibitem{Heyl15} M. Heyl, Phys. Rev. Lett. {\bf 115}, 140602 (2015).
\bibitem{Dora15} S. Vajna and B. Dora, Phys. Rev. B {\bf 91}, 155127 (2015).
\bibitem{BH16} J. C. Budich and M. Heyl, Phys. Rev. B {\bf 93}, 085416 (2016).
\bibitem{Balatsky} Z. Huang and A. V. Balatsky, Phys. Rev. Lett. {\bf 117}, 086802 (2016).
\bibitem{Budich16} Y. Hu, P. Zoller, and J. C. Budich, Phys. Rev. Lett. {\bf 117}, 126803 (2016).
\bibitem{Refael16} J. H. Wilson, J. C. W. Song, and G. Refael, Phys. Rev. Lett. {\bf 117}, 235302 (2016).
\bibitem{Bhaseen16} M. D. Caio, N. R. Cooper, and M. J. Bhaseen, Phys. Rev. B {\bf 94}, 155104 (2016).
\bibitem{Sondhi17} R. Moessner and S. L. Sondhi, Nat. Phys. {\bf 13}, 44 (2017).
\bibitem{Zhai17} C. Wang, P. Zhang, X. Chen, J. Yu, and H. Zhai, Phys. Rev. Lett. {\bf 118}, 185701 (2017).
\bibitem{Chen17} C. Yang, L. Li, and S. Chen, Phys. Rev. B {\bf 97}, 060304(R) (2018).
\bibitem{Ueda17} Z. Gong and M. Ueda, arXiv:1710.05289.
\bibitem{ETHcoldatom14} G. Jotzu, M. Messer, R. Desbuquois, M. Lebrat, T. Uehlinger, D. Greif, and T. Esslinger, Nature {\bf 515}, 237 (2014).
\bibitem{Jo17} B. Song, L. Zhang, C. He, T. F. J. Poon, E. Haiiyev, S. Zhang, X.-J. Liu, and G.-B. Jo, Science Advances {\bf 4}, 4748 (2018).
\bibitem{iondtop} P. Jurcevic, H. Shen, P. Hauke, C. Maier, T. Brydges, C. Hempel, B. P. Lanyon, M. Heyl, R. Blatt, and C. F. Roos, Phys. Rev. Lett. {\bf 119}, 080501 (2017).
\bibitem{Weitenberg17} N. Fl\"{a}schner, D. Vogel, M. Tarnowski, B. S. Rem, D.-S. L\"{u}hmann, M. Heyl, J. C. Budich, L. Mathey, K. Sengstock, and C. Weitenberg, Nat. Phys. {\bf 14}, 265 (2018)
\bibitem{Weitenberg1709} M. Tarnowski, F. Nur-Unal, N. Flaschner, B. S. Rem, A. Eckard, K. Sengstock, and C. Weitenberg, arXiv:1709.01046.
\bibitem{Xiong-Jun} L. Zhang, L. Zhang, S. Niu, X. -J. Liu, arXiv: 1802.10061.
\bibitem{shuaichen} W. Sun, C.-R. Yi, B.-Z. Wang, W.-W. Zhang, B. C. Sanders, X.-T. Xu, Z.-Y. Wang, J. Schmiedmayer, Y. Deng, X.-J. Liu, S. Chen, and J.-W. Pan, arXiv:1804.08226.
\bibitem{loschmidt} K. Brandner, V. F. Maisi, J. P. Pekola, J. P. Garrahan, and C. Flindt, Phys. Rev. Lett. {\bf 118}, 180601 (2017).

\bibitem{fisherbook} M. E. Fisher, The Nature of Critical Points, {\it Lectures in Theoretical Physics} Vol. 7 (University of Colorado Press, Boulder, CO, 1965).
\bibitem{LY52} C. N. Yang and T. D. Lee, Phys. Rev. {\bf 87}, 404 (1952); T. D. Lee and C. N. Yang, Phys. Rev. {\bf 87}, 410 (1952).


\bibitem{kehrein15} M. Schmitt and S. Kehrein, Phys. Rev. B {\bf 92}, 075114 (2015).
\bibitem{dutta17} U. Bhattacharya and A. Dutta, Phys. Rev. B {\bf 95}, 184307 (2017).
\bibitem{dutta17ti}  U. Bhattacharya and A. Dutta, Phys. Rev. B {\bf 96}, 014302 (2017).
\bibitem{higher} E. Canovi, P. Werner, and M. Eckstein, Phys. Rev. Lett. {\bf 113}, 265702 (2014).

\bibitem{sun16} J. Gu and K. Sun, Phys. Rev. B {\bf 94}, 125111 (2016).
\bibitem{hopf} T. Frankel, {\it The geometry of physics: an introduction} (Cambridge University Press, 2011).

\bibitem{haldane} F. D. M. Haldane, Phys. Rev. Lett. {\bf 61}, 2015 (1988).
\bibitem{qah1} Z. Wu, L. Zhang, W. Sun, X.-T. Xu, B.-Z. Wang, S.-C. Ji, Y. Deng, S. Chen, X.-J. Liu, and J.-W. Pan, Science {\bf 354}, 83 (2016).
\bibitem{qah2} W. Sun, B.-Z. Wang, X.-T. Xu, C.-R. Yi, L. Zhang, Z. Wu, Y. Deng, X.-J. Liu, S. Chen, and J.-W. Pan, arXiv:1710.00717.

\end{thebibliography}
\end{document}